\nonstopmode

\def\ket#1{\left|#1\right\rangle}

\documentclass[aps,twocolumn,superscriptaddress]{revtex4}
\usepackage{graphicx,amsmath,bbm}

\begin{document}
\title{Dynamics of spin-flip photon-assisted tunneling}
\author{F.~R. Braakman}
\affiliation{Kavli Institute of Nanoscience, TU Delft, 2600 GA Delft, The Netherlands}
\affiliation{Department of Physics, University of Basel, 4056 Basel, Switzerland}
\author{J. Danon}
\affiliation{Dahlem Center for Complex Quantum Systems, Freie Universit\"at Berlin, Germany}
\affiliation{Niels Bohr International Academy, Niels Bohr Institute, University of Copenhagen, Denmark}
\author{L.~R. Schreiber}
\affiliation{Kavli Institute of Nanoscience, TU Delft, 2600 GA Delft, The Netherlands}
\affiliation{II. Institute of Physics, RWTH Aachen University, 52074 Aachen, Germany}
\author{W. Wegscheider}
\affiliation{Solid State Physics Laboratory, ETH Z\"{u}rich, 8093 Z\"{u}rich, Switzerland}
\author{L.~M.~K. Vandersypen}
\affiliation{Kavli Institute of Nanoscience, TU Delft, 2600 GA Delft, The Netherlands}
\date{\today}

\begin{abstract}
We present time-resolved measurements of spin-flip photon-assisted tunneling and spin-flip relaxation in a doubly occupied double quantum dot. The photon-assisted excitation rate as a function of magnetic field indicates that spin-orbit coupling is the dominant mechanism behind the spin-flip under the present conditions. We are able to extract the resulting effective `spin-flip tunneling' energy, which is found to be three orders of magnitude smaller than the regular spin-conserving tunneling energy. We also measure the relaxation and dephasing times of a qubit formed out of two two-electron states with different spin and charge configurations.
\end{abstract}

\maketitle
\section{Introduction}
\label{sec:intro}
The manipulation of electron spins in gated double quantum dots (DQDs) forms a topic of intense research in recent years, particularly in the context of quantum information~\cite{Loss98,Hanson07,Taylor05}. Of special interest in this field is the case of a DQD occupied with two electrons~\cite{Levy02,Ono02}. Transitions between the spin states of the two electrons have been realized using the exchange interaction~\cite{Petta05} or Landau-Zener processes~\cite{Petta10}. Interestingly, the spin of an electron can also be flipped by letting the electron tunnel from one dot to another in the presence of a magnetic field gradient or finite spin-orbit interaction. In recent work~\cite{Schreiber11}, we demonstrated coupling of different two-electron spin states by photon-assisted tunneling~\cite{Wiel03} (spin-flip PAT). A theoretical estimate of the strength of the various spin-flip tunneling mechanisms pointed in the direction of spin-orbit interaction as being dominant. Furthermore, it was theoretically shown that this process in principle enables full coherent manipulation and read-out in the two-spin manifold.

In this paper, we experimentally probe and theoretically analyze time-resolved spin-flip PAT transitions. We focus on two states, a spin singlet with both electrons occupying the right dot and a spin triplet with one electron in each dot, and we investigate the dynamics of microwave-induced transitions as well as phonon-mediated relaxation between these states. In the regime of our measurements (with an applied external magnetic field exceeding 1 T) we can confirm the prominent role of spin-orbit interaction for the spin-flip tunnel coupling. For our specific geometry, we find the resulting effective photon-assisted coupling energy to be $\sim 10^{-8}$ eV (depending on the microwave power applied), which would correspond to Rabi oscillations on a MHz timescale. Furthermore, from the measurements we can extract the relaxation and dephasing times $T_1$ and $T^*_2$ for the two-level system, yielding $T_1 = 14~\mu$s (at a level splitting of 85 $\mu$eV) and $T^*_2 = 280$ ps. Given the present decay times, no Rabi oscillations are visible in the data and the observed dynamics of the transitions are mostly incoherent.

\section{Setup and spin-flip PAT}
\begin{figure*}[t]
\centering
\includegraphics[scale=0.75]{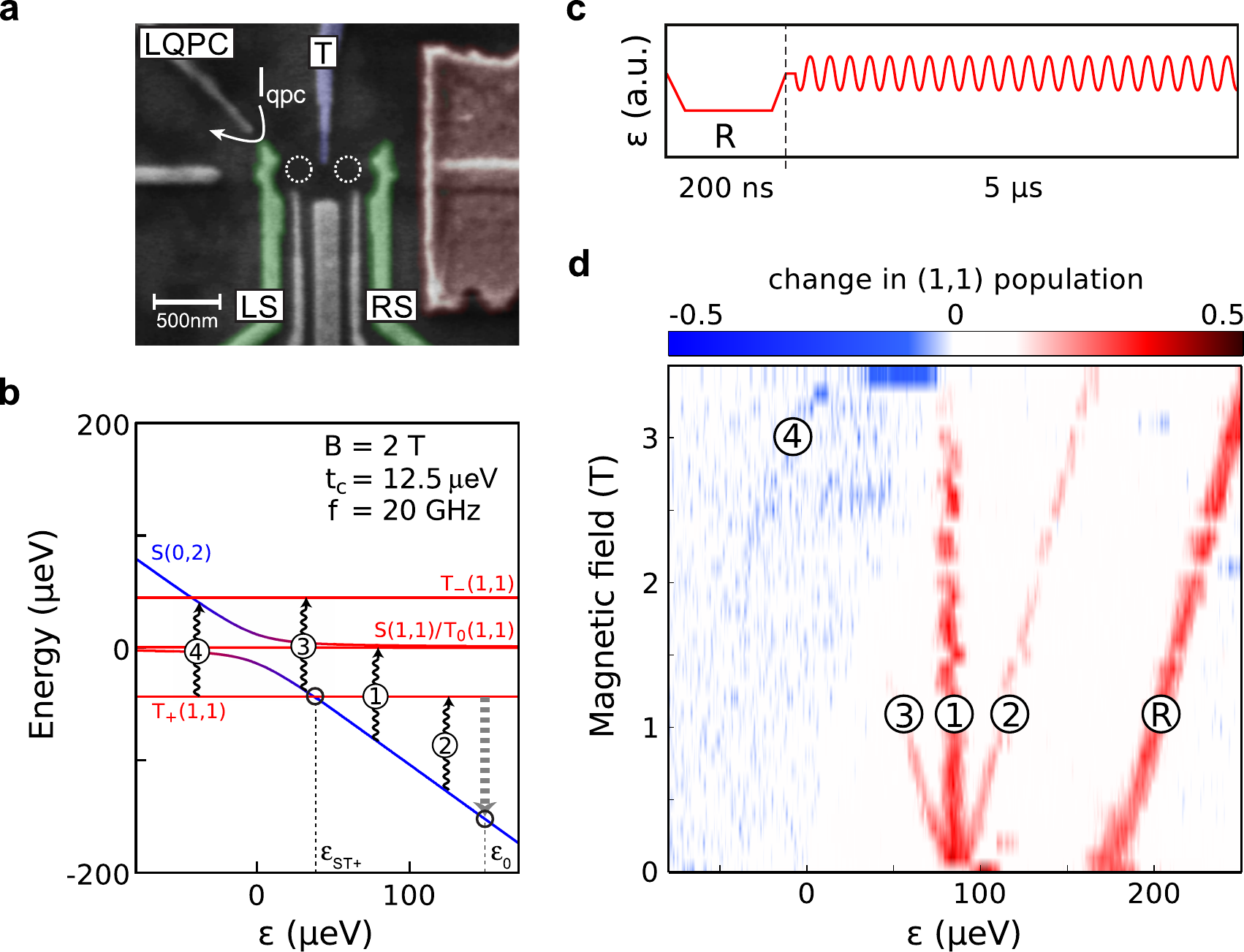}
\caption{(color online) (a) False color SEM image of a sample similar to the one used here. Dotted circles indicate the approximate location of the DQD. Current flow through the QPC is represented by the white arrow. The micromagnet is colored red. (b) Schematic energy level diagram of the doubly occupied DQD. Black wavy arrows show the possible PAT transitions. The gray dashed arrow represents the relaxation process $T_+\to S(0,2)$ investigated in Sec.~\ref{sec:rel}. (c) Sketch of the pulsing scheme used to obtain the PAT spectrum shown in (d): Continuous microwaves are applied to RS, interspersed with short pulses amounting to shifts along the detuning axis. (d) Measurement of the QPC lock-in signal, normalized to represent the change in population of the $(1,1)$ charge configuration (see Sec.~\ref{sec:rel} for the normalization procedure). The peaks labeled 1 to 4 correspond respectively to the photon assisted transitions $S(0,2) \to S(1,1)$, $S(0,2) \to T_+$, $S(0,2) \to T_-$, and $T_+ \to S$.}
\label{fig:spatiifig1}
\end{figure*}

Fig.~\ref{fig:spatiifig1}a shows a scanning electron micrograph of a device similar to the one used in the experiment. A two-dimensional electron gas (2DEG) is formed 90 nm below the surface of the GaAs/AlGaAs heterostructure, in which a DQD is created by the application of negative bias voltages to the metallic (Ti-Au) gates defined on the surface. The DQD interdot axis is aligned along the $[110]$ GaAs crystallographic direction. The device is mounted inside a dilution refrigerator with a base temperature of 30 mK. 

The voltages on the various gates can be used to tune different properties of the DQD. The gate settings are chosen such that the DQD behaves as a relatively closed system: The tunneling rates from the DQD to the leads are $> 10^{3}$ times smaller than the interdot tunneling rate. Gates LS and RS are further used to set the electrochemical potential of the dots relative to the Fermi level in the contacting reservoirs, enabling control over the charge occupation of both dots. Gate T is used to set the tunnel barrier between the dots. Read-out of the occupation of each dot is performed through charge sensing with a quantum point contact~\cite{Field93} (QPC), formed in the 2DEG by gates LS and LQPC. The conductance of the QPC is tuned to be on the flank of the first conductance plateau, where it is very sensitive to changes in the electrostatic potential, permitting read-out of the number of charges on each quantum dot. In addition, high-frequency voltage excitations in the form of microwaves and pulses can be applied via bias-tees to gates LS and RS. In all measurements presented here, standard lock-in techniques were used to read out the conductance of the QPC, while the microwave and pulsed excitations were chopped at the reference frequency of the lock-in amplifier (380 Hz). In this way, the measured lock-in signal reflects the change in the double-dot occupation due to the excitations. A magnetic field of magnitude $B$ was applied along the DQD interdot-axis. Finally, a cobalt micromagnet (red rectangle) was also present in the system, creating a magnetic field gradient across the dots. Although this field gradient in principle also provides a mechanism for spin-flip tunneling between different two-electron states, the strength of the coupling was estimated to be much lower than that provided by the spin-orbit interaction~\cite{Schreiber11}. 

\begin{figure}[t]
\centering
\includegraphics[scale=0.75]{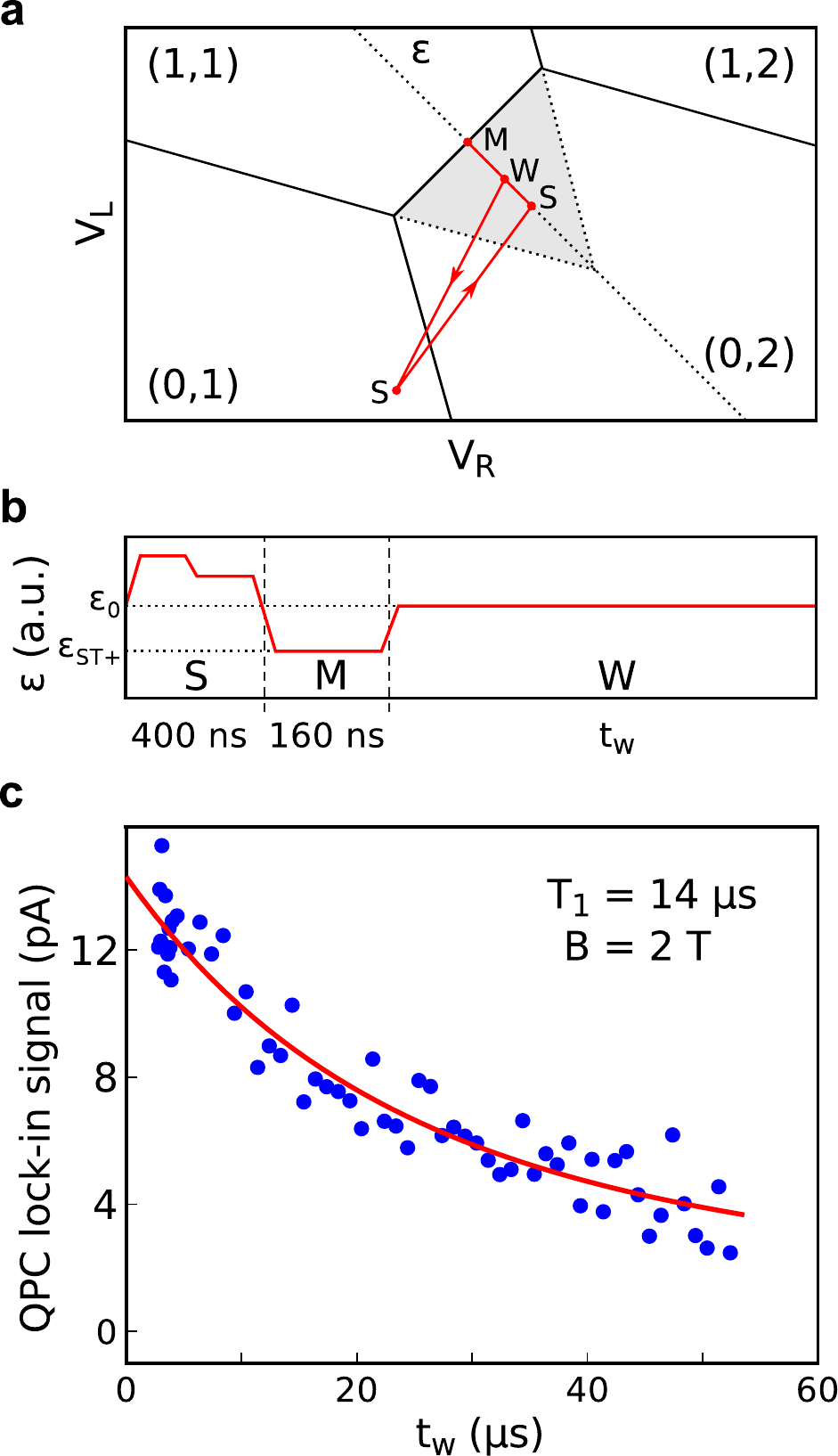}
\caption{(color online) (a) Schematic showing the excitation cycle used in the measurements of $T_1$, drawn in the charge stability diagram of the DQD. The dotted diagonal line indicates the detuning axis $\varepsilon$. (b) The projection of this cycle on the detuning axis. (c) QPC lock-in signal as function of the waiting time $t_w$. The red curve is a fit to an exponential decay, averaged over the cycle time.}
\label{fig:spatiifig2}
\end{figure}

The DQD is tuned to the two-electron regime where the relevant two-electron states are $S(0,2)$, $S(1,1)$, $T_-(1,1)$, $T_0(1,1)$ and $T_+(1,1)$. Here $S$ denotes a spin singlet state and $T_{-,0,+}$ spin triplet states corresponding to a value of $m_s = -1,0,1$ respectively, with $m_s$ being the quantum number for the spin projection along the direction of the external magnetic field. The bracketed numbers indicate the charge occupation of the left and right dot respectively. In Fig.~\ref{fig:spatiifig1}b we sketch the energies of the five states as a function of the detuning $\varepsilon$ between the electrochemical potentials of the two singlet states $S(0,2)$ and $S(1,1)$ (here $\varepsilon$ is defined to correspond to an equal but opposite change in the energies of the $(0,2)$ and $(1,1)$ states, i.e.\ it is the axis perpendicular to the $(0,2)$-$(1,1)$ charge transition line, see Fig.\ \ref{fig:spatiifig2}a). In the plot the magnetic field is set to $B = 2$~T (using a $g$-factor of $g=0.38$ \cite{Schreiber11}) and the interdot tunnel coupling to $t_c=12.5$ $\mu$eV. The energy splitting between $S(0,2)$ and the various $(1,1)$ states can be made resonant with the frequency of an applied microwave field by varying the detuning $\varepsilon$ or the magnetic field $B$. In the Figure, four of these resonances are indicated by black wavy arrows (for a microwave frequency of $f$ = 20 GHz). In the presence of a spin-flipping mechanism, all these resonances could lead to photon-assisted charge transitions, which in principle could be detected by the QPC charge sensor.

Indeed, when we apply microwaves to RS (periodically interspersed with short gate-voltage pulses, as shown in Fig.~\ref{fig:spatiifig1}c), several resonances appear in the QPC signal, see Fig.~\ref{fig:spatiifig1}d. The data presented in Fig.~\ref{fig:spatiifig1}d and their interpretation have already been published in Ref.~\cite{Schreiber11}, but are included to make the paper self-contained. Fig.~\ref{fig:spatiifig1}d shows the lock-in signal of the charge sensor as a function of detuning and magnetic field. This signal has been normalized to represent the average change in the charge occupation of the DQD induced by the presence of the microwave excitation and voltage pulses (see Sec.~\ref{sec:rel}).
The voltage pulses (`R' in Fig.~\ref{fig:spatiifig1}c) result in the line labeled `R' in Fig.~\ref{fig:spatiifig1}d, which serves as a reference along the detuning axis, as explained in \cite{Schreiber11}. The other four observed resonances can be identified with the PAT processes indicated in Fig.~\ref{fig:spatiifig1}b by the wavy arrows. The line labeled 1 does not move with magnetic field, which identifies it as the spin-conserving $S(0,2)\to S(1,1)$ transition (the $S(0,2)$-$T_0$ resonance, which is indistinguishably close in detuning, does not contribute significantly to the observed charge transitions, see \cite{Schreiber11}). The lines 2 and 3 move to lower and higher detunings respectively with increasing magnetic field, labeling them the $S(0,2) \to T_-$ and $S(0,2)\to T_+$ transitions. Line 4, the faint blue line in the upper left corner of the plot, corresponds to the transition $T_+ \to S$, where $S$ is a hybridization of $S(0,2)$ and $S(1,1)$, see Fig.~\ref{fig:spatiifig1}a. The bending of this line directly reflects the curvature of the avoided crossing of the $(0,2)$ and $(1,1)$ singlet states around zero detuning, created by the tunnel coupling. This thus allows for a direct determination of the interdot tunnel coupling \cite{Schreiber11}.

The use of microwave bursts and voltage pulses makes it possible to investigate the dynamics of the photon-assisted mixing processes observed in Fig.~\ref{fig:spatiifig1}d, as well as the relaxation rate from an excited state to the ground state. Here, we will focus on mixing and relaxation in the $S(0,2)$-$T_+$ subspace (the transition labeled `2' in Fig.~\ref{fig:spatiifig1}): this enables us to study the spin-flip mechanism in detail, while simultaneously having high signal amplitudes at all values of the magnetic field that were used. Similar analyses apply to the other spin-flip transitions. 

\section{Spin-flip relaxation rate}\label{sec:rel}
First, we investigate the spin-flip relaxation rate from an excited $T_+(1,1)$ state to the $S(0,2)$ ground state. We apply a magnetic field of 2 T, at which spin-flip relaxation is predicted to be mainly mediated by spin-orbit coupling, accompanied with the emission of a phonon~\cite{Danon13,Hanson07}.

We use a pulsing scheme as shown in Fig.~\ref{fig:spatiifig2}a and b. Each cycle starts with a `reset' pulse (`S' in Fig.~\ref{fig:spatiifig2}a), which involves pulsing to the $(0,1)$ charge configuration and then into the $(0,2)$ regime, in such a way that at the end of `S' the DQD is in the $S(0,2)$ ground state. The system is then pulsed to the $S$-$T_+$ mixing point (denoted by $\varepsilon_{ST_+}$ in Fig.~\ref{fig:spatiifig1}b) where the energies of the hybridized singlet $S(0,2)/S(1,1)$ and the triplet $T_+(1,1)$ states cross. These states are coupled by spin-orbit interaction as well as by small gradients in the $x$- and $y$- components of the magnetic field, originating from differences in the Overhauser fields in both dots~\cite{Hanson07} and possibly from the micromagnet. The detuning is kept at $\varepsilon_{ST_+}$ for 160 ns (`M' in Fig.~\ref{fig:spatiifig2}a), which is long enough to obtain complete mixing between the singlet and triplet state. Next, the system is brought to the point $\varepsilon_0 = \varepsilon_{ST_+} + \varepsilon_\Delta$, where we wait for a variable waiting time $t_w$ (`W' in Fig.~\ref{fig:spatiifig2}a). Then the cycle is repeated.

Directly after pulsing from $\varepsilon_{ST_+}$ to $\varepsilon_0$, in 50\% of the cycles the system will be measured to be in the $T_+(1,1)$ state, from which relaxation to $S(0,2)$ is suppressed by Pauli spin blockade~\cite{Ono02}. Since the QPC distinguishes between the two charge configurations, the lock-in signal is proportional to the population of the $T_+(1,1)$ state, averaged over the cycle time. In Fig.~\ref{fig:spatiifig2}c we plot the amplitude of the QPC lock-in signal as a function of $t_w$ (blue dots), and we indeed observe a decaying trend corresponding to the (slow) relaxation from $T_+$ to $S(0,2)$ (indicated by the thick dashed gray arrow in Fig.~\ref{fig:spatiifig1}b). From these data we can extract the relaxation time $T_1$ by fitting the QPC signal to a time-averaged exponential decay function, $I_{QPC} \propto (1/t_w) \int_{0}^{t_w} \exp(-t/T_1)dt$. We thus neglect the duration of the parts `S' and `M' of the pulse cycle compared to $t_w$. The fit resulted in $T_1=14$ $\mu$s at $B=2$ T and $\varepsilon_\Delta = 85~\mu$eV.

If we extrapolate the relaxation curve to $t_w = 0$, we can read off the QPC lock-in signal which corresponds to a 50\% occupation probability of the $(1,1)$ charge state. We find this signal to be 14.3 pA. Of course, this normalization factor has been determined for a very specific point in gate space (where $\varepsilon \approx g\mu_{\rm B} B + \varepsilon_\Delta \approx 130~\mu$eV). However, when we varied $\varepsilon_\Delta$, we did not observe a significant change in the estimated normalization factor.

\section{Dynamics of spin-flip PAT}

\begin{figure}[t]
\centering
\includegraphics[scale=0.75]{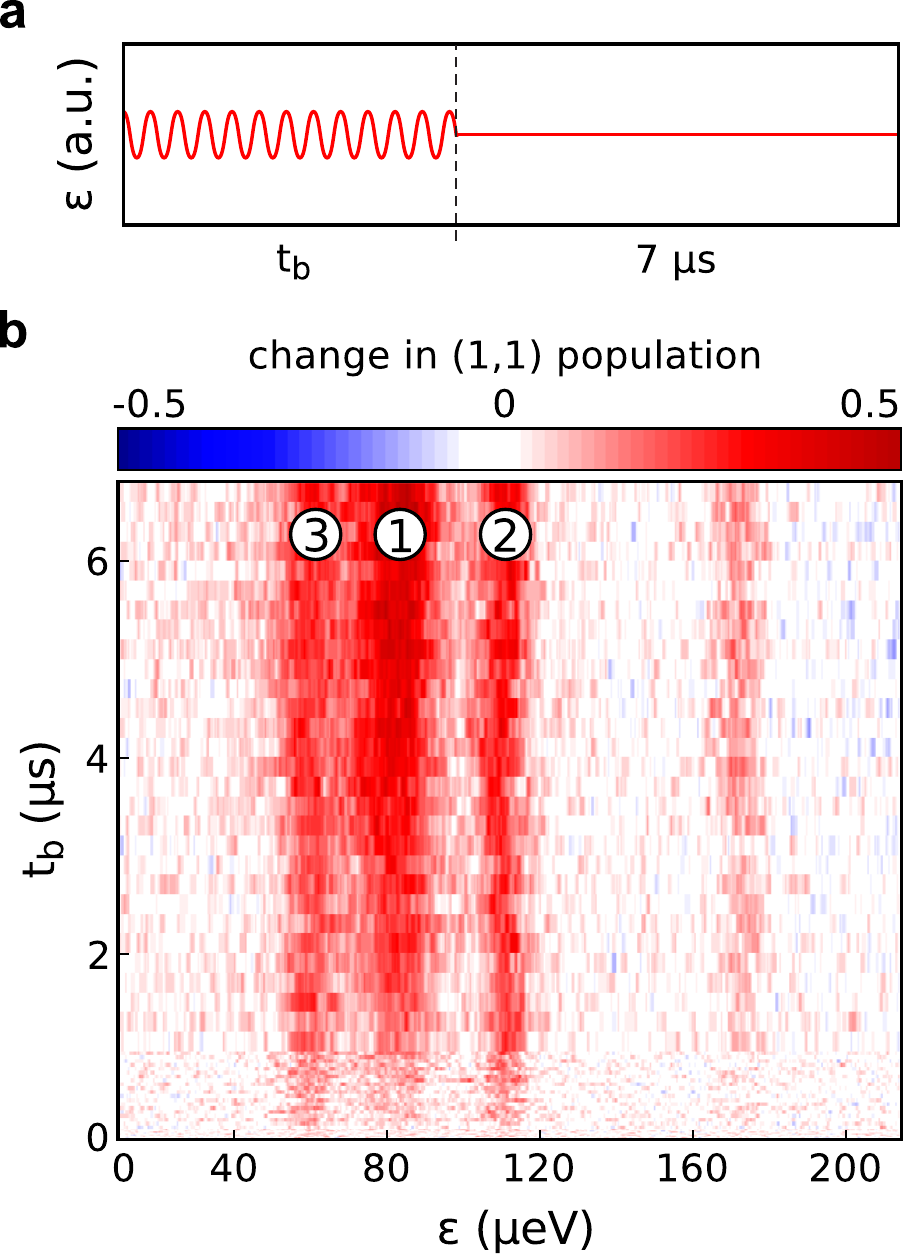}
\caption{(color online) (a) Schematic showing the excitation cycle used in the time-resolved spin-flip PAT measurements. (b) QPC lock-in signal as function of detuning $\varepsilon$ and burst time $t_b$. We applied a magnetic field of 1 T and the power of the microwave source was set to 12 dBm.}
\label{fig:spatiifig3}
\end{figure}
Now we examine the buildup of the populations of the $T_+(1,1)$ state during microwave excitation in order to extract the microwave-driven coupling energy between $S(0,2)$ and $T_+$, as well as the dephasing time $T_2^*$. The pulse cycle is schematically shown in Fig.~\ref{fig:spatiifig3}a: A microwave signal with a frequency $f=20$~GHz is applied during a time $t_b$, after which the detuning is kept constant for 7~$\mu$s to allow for (partial) relaxation back to the $S(0,2)$ ground state. In Fig.~\ref{fig:spatiifig3}b we show the resulting normalized QPC lock-in signal as a function of detuning $\varepsilon$ and burst time $t_b$. Here, the magnetic field was set to 1 T and the power of the microwave source was 12 dBm. When scanning along the detuning axis, we can distinguish four different resonances: We observe three peaks at $\varepsilon \approx 60$, 85, and 110 $\mu$eV, corresponding to the $S(0,2) \to T_-$, $S(0,2) \to S(1,1)$, and $S(0,2) \to T_+$ transitions (labeled respectively `3', `1', and `2' in Fig.~\ref{fig:spatiifig1}). Besides, we see a faint peak at $\varepsilon \approx 170~\mu$eV, which we identify as the 2-photon transition from $S(0,2)$ to $S(1,1)$.

We focus on the $S(0,2) \to T_+$ transition (third peak from the left). A first observation is that the time-dependent signal does not seem to exhibit Rabi oscillations. Since the two states involved in the driving have different charge configurations, the overall coherence is dominated by the relatively short charge coherence time, which is typically sub-nanoseconds~\cite{Petersson10}. The main source for this fast decoherence is believed to be $\sim 1/f$-noise coupling to the qubit splitting, caused by slowly fluctuating charges in the environment~\cite{Yurkevich10}. Indeed, experiments probing the spectrum of the charge fluctuations in the electrostatic environment of a DQD, yield a $1/f^{0.7}$-divergence at small frequencies~\cite{Dial13}. Given the short expected coherence times and the absence of Rabi oscillations in our signal, we assume in what follows that the microwave-induced driving rate is slow compared to the dephasing timescale.

To model the dynamics of the $S(0,2)$-$T_+$ two-level system during the excitation pulse, we write a Hamiltonian
\begin{equation}
\hat H = \frac{1}{2} \big( E_z + \tilde \varepsilon \cos \omega t \big)\hat \sigma_z + q \hat \sigma_x,
\end{equation}
where $E_z$ is the level splitting between $S(0,2)$ and $T_+$, $\omega = 2\pi f$ is the microwave driving frequency, $\tilde \varepsilon$ is the driving amplitude along the detuning axis of the DQD, and $q$ represents the coupling energy between $S(0,2)$ and $T_+$. The operators $\hat{\boldsymbol\sigma}$ are the Pauli matrices, acting in the two-dimensional space $\{ \ket{T_+}, \ket{S(0,2)} \}$.

We then apply a unitary transformation to a non-uniformly rotating frame, $\hat H' = e^{-\frac{i}{2}\phi(t)\hat\sigma_z}\hat H e^{\frac{i}{2}\phi(t)\hat\sigma_z}$ with $\phi(t) = \frac{1}{\hbar}\int_0^t d\tau\, \tilde\varepsilon \cos \omega \tau = (\tilde\varepsilon / \hbar\omega)\sin\omega t$~\cite{Oliver05}. This yields
\begin{equation}
\hat H' = \frac{1}{2} E_z \hat \sigma_z + q \sum_n J_n(\alpha) \big[ e^{-in\omega t}\hat\sigma^+ +e^{in\omega t}\hat\sigma^-\big],
\end{equation}
where $J_n(x)$ is the $n$-th Bessel function of the first kind and $\alpha = \tilde\varepsilon / \hbar\omega$ is the normalized driving amplitude. Assuming that $q \ll E_z$ and that $E_z \approx \hbar\omega$ we focus on the 1-photon term and write $\hat H'$ again in a rotating frame,
\begin{equation}
\hat H_{\rm eff} = \frac{1}{2} \Delta \hat \sigma_z +  q J_1(\alpha) \hat \sigma_x,\label{eq:heff}
\end{equation}
where $\Delta = E_z - \hbar\omega$ is the detuning from the 1-photon resonance.

We then determine the time-dependence of the density matrix $\hat \rho$ describing the two-level system. We first investigate the limit where coupling between $S(0,2)$ and $T_+$ is absent ($q\to 0$) and thus also no phonon-mediated relaxation is present. We model the $1/f$ charge fluctuations as quasistatic Gaussian fluctuations that are added to $\Delta$ (static on the timescale of a single pulse cycle). We then find the diagonal elements to be constant in time, and for the off-diagonal elements this results in
\begin{equation}
\rho_{+S}(t) = [\rho_{S+}(t)]^* = \rho_{+S}(0) e^{-\frac{i}{\hbar}\Delta t}e^{-\frac{1}{2}\sigma^2t^2},\label{eq:od}
\end{equation}
where the index $+$ refers to $T_+$. The first exponential in (\ref{eq:od}) describes coherent precession of the state vector on the Bloch sphere driven by the energy splitting $\Delta$. The second exponential describes the dephasing of the density matrix in the limit of quasi-static (classical) fluctuations coupling to $\hat \sigma_z$. Indeed, such fluctuations are known to result in `Gaussian' dephasing on a timescale of $T_2^* = \sqrt{2}/\sigma$, where $\hbar^2\sigma^2 = \langle \delta \varepsilon^2 \rangle$ is the variance of the fluctuations along $\hat\sigma_z$~\cite{Ithier05}. As an extra test of our assumptions, we also performed the theoretical analysis presented in this Section under the opposite assumption of white charge noise, leading to exponential dephasing $\propto e^{-t/T_2^*}$. Given the physics of charge noise, this assumption is much less plausible, and it indeed resulted in a less consistent set of results~\cite{footnote2}.

\begin{figure*}[t]
\centering
\includegraphics[scale=0.75]{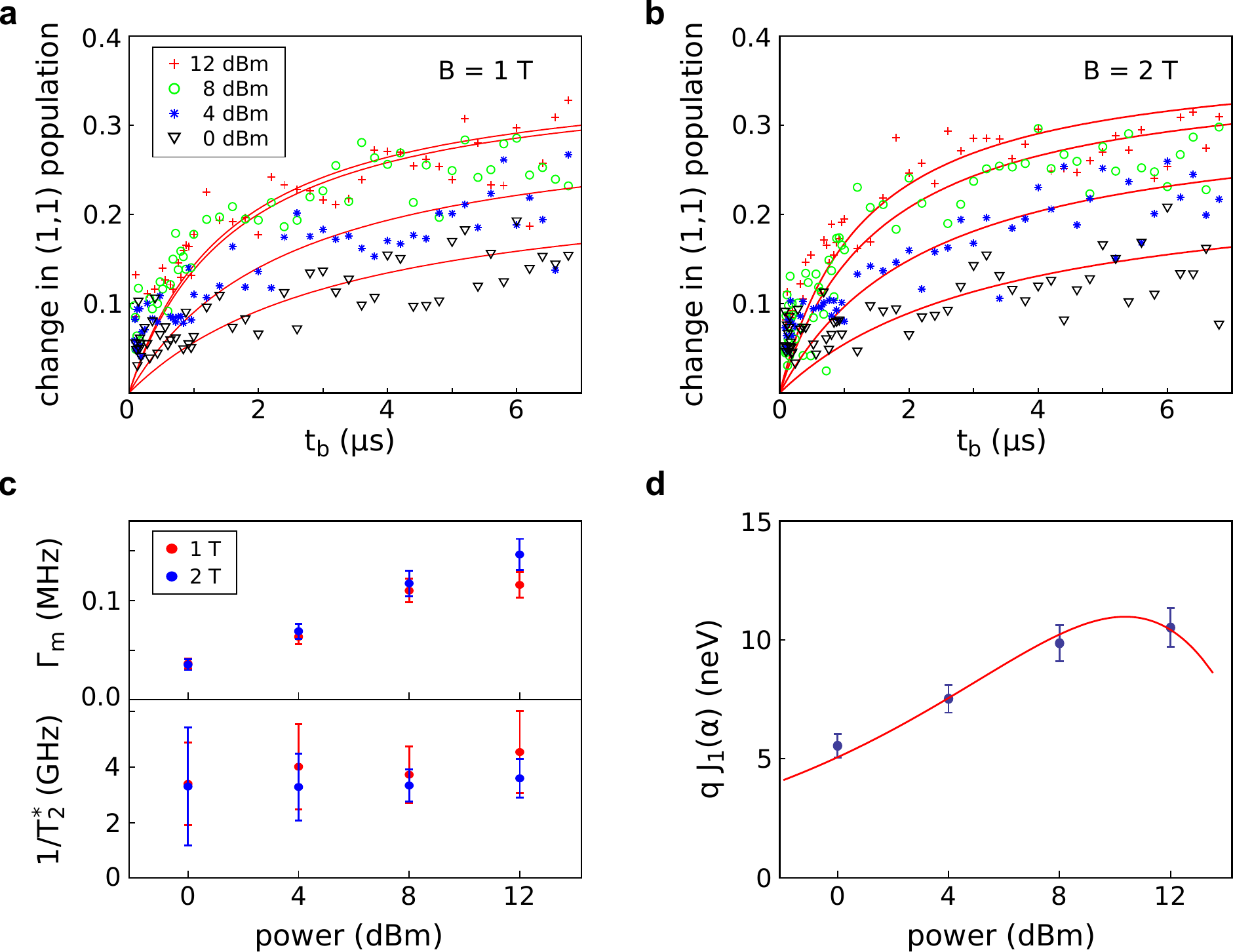}
\caption{(color online) (a,b) Measured population of the $(1,1)$ charge state as a function of microwave burst time $t_b$, for different microwave powers and magnetic fields (all for the $S$-$T_+$ resonance), averaged over the measurement cycle. The attenuation of the high-frequency wiring was measured to be 66 dB at 20 GHz; for an output of 0 dBm power of the microwave source, this corresponds to a voltage of 230 $\mu$V on the open-ended gate. The red curves are fits used to extract $\Gamma_m(0)$. (c) Extracted `mixing rates' $\Gamma_m(0)$ and decoherence rates $1/T_2^* = \sigma/\sqrt{2}$ from the data shown in (a,b). (d) Blue dots: The effective coupling elements $qJ_1(\alpha)$ found from the values of $\Gamma_m(0)$ and $T_2^*$ presented in (c). Red curve: A fit to $q J_1(\alpha_0 10^{P/20})$, where $P$ is the applied microwave power at the source in dBm.}
\label{fig:spatiifig4}
\end{figure*}
The coupling element $q$ is then added as a small perturbation. To find a time-evolution equation for the density matrix, we employ a second order perturbation theory assuming fast dephasing, $\hbar\sigma \gg q$. Adding a phenomenological rate $\Gamma$ accounting for the phonon-mediated relaxation, we then find for the diagonal elements of the density matrix
\begin{align}
\frac{d\rho_+}{dt} & = \Gamma_m(\Delta) \rho_S - [\Gamma_m(\Delta) + \Gamma] \rho_+, \label{eq:drdt1} \\
\frac{d\rho_S}{dt} & = [\Gamma_m(\Delta) + \Gamma] \rho_+ - \Gamma_m(\Delta) \rho_S, \label{eq:drdt2}
\end{align}
where $\Gamma_m(\Delta) = [\sqrt{2\pi}q^2 J_1(\alpha)^2/\hbar^2\sigma] \exp\{ -\Delta^2 / 2\hbar^2\sigma^2 \}$ is the transition rate due to the microwave field. The dynamics of the off-diagonal elements are still dominated by dephasing on the timescale $T_2^* \ll \Gamma_m^{-1}, \Gamma^{-1}$, so coherence can be neglected and we can focus on the master equations (\ref{eq:drdt1}) and (\ref{eq:drdt2}).

With the initial condition $\rho_+(t=0)=0$, the solution of (\ref{eq:drdt1}) and (\ref{eq:drdt2}) reads
\begin{equation} 
\rho_+(\Delta,t) = \frac{\Gamma_m(\Delta)}{\Gamma+2\Gamma_m(\Delta)} \big\{ 1-e^{-[\Gamma+2\Gamma_m(\Delta)]t} \big\}.
\label{eq:tdsol}
\end{equation}
In steady state, i.e.\ for large times $t > \Gamma_m^{-1}, \Gamma^{-1}$, the line shape approaches the non-Lorentzian form
\begin{equation}
\rho_+(\Delta) = \frac{\Gamma_m(\Delta)}{\Gamma + 2\Gamma_m(\Delta)}  = \frac{\gamma}{2\gamma + \exp \{\frac{\Delta^2}{2\hbar^2\sigma^2} \}},\label{eq:gauss}
\end{equation}
with $\gamma \equiv \Gamma_m(0) / \Gamma$ characterizing the strength of the driving compared to the relaxation rate. We see that $0<\rho_+<\tfrac{1}{2}$ depending on $\gamma$ and that the line width FWHM $= 2\hbar\sigma \sqrt{2 \ln (2\gamma + 2)}$ scales with $\sigma$ and becomes power-broadened with increasing $\gamma$. We note that, in contrast to a Lorentzian curve, here the line {\it shape} also changes with power broadening: (i) For $\gamma \ll 1$, it follows that $\Gamma \gg \Gamma_m(\Delta)$ for all $\Delta$, i.e.\ microwave driven transitions are rare and suppressed everywhere. Steady state results in a faint response centered at $\Delta=0$, broadened by the Gaussian noise. This yields a Gaussian peak $\rho_+(\Delta) \approx \Gamma_m(\Delta)/\Gamma$. (ii) For $\gamma \gg 1$ relaxation is slow compared to driving, and $\rho_+ \approx \tfrac{1}{2}$ as long as $\Gamma$ can be neglected, which is true for a broad range of $\Delta$ (the condition is $\Delta^2 < 8\hbar^2\sigma^2 \ln 2\gamma$). In this case the peak has a broad flat top.

We use the line shape given in (\ref{eq:gauss}) to fit data such as shown in Fig.~\ref{fig:spatiifig3}b: We first subtract a smooth background signal (to compensate for capacitive coupling between the gates used to sweep the detuning and the QPC) and then fit for each different $t_b$ the measured QPC lock-in signal as a function of $\varepsilon$ to two, three, or four curves, depending on how many resonances are visible. In Fig.~\ref{fig:spatiifig4}a,b we show the resulting fitted peak heights for the $S$-$T_+$ resonance, normalized to 50\% $=$ 14.3 pA, as a function of $t_b$ for different microwave powers and two different magnetic fields.

These time-dependent data are then fitted to the averaged charge population calculated for a continuously repeated cycle as shown in Fig.~\ref{fig:spatiifig3}a. We use $\Gamma = 14~\mu$s, as we found above for a level splitting of $85~\mu\text{eV} \approx hf$, which allows us to extract $\Gamma_m(0)$ for each curve. For larger burst times the resonance approaches the Gaussian form given in (\ref{eq:gauss}). Since the measured line width is found to be almost constant for $t_b > 3~\mu$s, we can use the fitted FWHM's in this regime to determine $\sigma$ as well. The resulting mixing rates $\Gamma_m(0)$ and decoherence rates $1/T_2^* = \sigma/\sqrt{2}$ are shown in Fig.~\ref{fig:spatiifig4}c. We make two observations: (i) $\Gamma_m(0) \propto q^2$ does not vary much between the data at 1 and 2 T. This is again a strong indication that spin-orbit coupling provides the dominant spin-mixing matrix elements. Indeed, processes involving a magnetic field gradient rely on mixing of the $T_+(1,1)$ and $S(1,1)$ states and tunnel coupling of $S(1,1)$ to $S(0,2)$, thus providing a small coupling between the $T_+$-like state and $S(0,2)$. The degree of mixing by the gradient is to first order $\sim \delta B / B$, where $\delta B$ is the gradient. For the resulting gradient-induced mixing rates we thus expect $\Gamma_m \propto q^2 \propto 1/B^2$. (ii) The decoherence rate $1/T_2^*$ seems to be fairly constant for different microwave powers and magnetic fields. This is consistent with the picture that background charge fluctuations provide the main dephasing mechanism: We expect that $\hbar^2\sigma^2 = \langle \delta \varepsilon^2 \rangle$, where the typical amplitude of the charge fluctuations $\langle \delta \varepsilon^2 \rangle^{1/2}$ is not influenced directly by the microwave power or $B$. From the eight data points shown in Fig.~\ref{fig:spatiifig4}c, we find $\langle \delta \varepsilon^2 \rangle^{1/2} = 3.28 \pm 0.36~\mu$eV or $T_2^* = 283 \pm 31$~ps.

Finally, we would like to estimate the magnitude of the spin-flip matrix element $q$ coupling $S(0,2)$ and $T_+$. From the values found for $\Gamma_m(0)$ and $\sigma$, we can calculate $q J_1(\alpha) = \hbar \sqrt{\Gamma_m(0) \sigma/\sqrt{2\pi}}$, the result is shown in Fig.~\ref{fig:spatiifig4}d. To arrive at a quantitative estimate for $q$, we then need to know $\alpha = \tilde\varepsilon/\hbar\omega$ as a function of the applied power $P$ (in dBm). The four data points are expected to fall on the curve $q J_1(\alpha_0 10^{P/20})$, where $\alpha_0 = \tilde\varepsilon(0)/\hbar\omega$ is the value for $\alpha$ at 0 dBm. Indeed, with this curve, a good fit can be produced (shown in red in the plot), which yields $q =  18.9 \pm 2.2$ neV with $\alpha_0 = 0.56 \pm 0.10$, which is consistent with our order-of-magnitude estimate for $\alpha_0$~\cite{footnote1}.

The spin-flip tunneling matrix element $q$ is thus found to be approximately a factor $1.5 \times 10^{-3}$ smaller than the `regular' tunneling matrix element $t \approx 12.5~\mu$eV. For the present geometry, this ratio can be estimated to be $|q/t| \approx d/\sqrt{2}l_{\rm so}^z$~\cite{Danon13}, where $d$ is the interdot distance and $l_{\rm so}^z$ is the spin-orbit length along the interdot axis. With $d=75$ nm and $l_{\rm so}^z \sim 10~\mu$m, we estimate $|q/t| \sim 5\times 10^{-3}$, which is of the same order of magnitude.

Placing these results in the context of quantum information, we compare the Rabi frequencies that are in principle obtainable by spin-flip PAT to other techniques of coherent spin manipulation. For the driving powers used here, we found $q J_1(\alpha) \approx 5$--$10$ neV, which would correspond to a coherent Rabi frequency of 1--2 MHz. This is slightly slower than the single-spin rotations achieved using electron spin resonance and spin-orbit mediated electric dipole spin resonance~\cite{Koppens06,Nowack07}, with Rabi frequencies of 5--10 MHz. Spin-flip PAT transitions between the other $(1,1)$ states and $S(0,2)$ should have similar Rabi frequencies, thus enabling in principle full coherent control of the $(1,1)$ manifold on similar timescales via Raman driving~\cite{Schreiber11}. A drawback of the spin-flip PAT investigated here is that the coherence times of the involved states, $T_2^* \approx 0.3$ ns is several orders of magnitude shorter than the Rabi period, making it presently not possible to observe coherent dynamics. We stress that these short coherence times stem from the sensitivity of the charge character of the spin-flip PAT transitions to slow fluctuations in the electrostatic environment of the DQD, and is not intrinsic to the spin-orbit mediated flipping mechanism. Also phonon absorption and emission is not expected to preclude coherent control of the $(1,1)$ manifold, see~\cite{Schreiber11}.

Summarizing, we have performed a time-resolved analysis of spin-flip PAT and spin-flip relaxation between the $T_+(1,1)$ and $S(0,2)$ two-electron states in a DQD. We confirmed the dominant role of spin-orbit interaction for the effective `spin-flip tunnel coupling' between these two states at higher magnetic fields ($B \geq 1$ T), and we were able to extract the corresponding coupling energy. Independently, we could determine the relaxation time $T_1$ and dephasing time $T_2^*$ for the two-level system formed by $T_+(1,1)$ and $S(0,2)$ at a level splitting of 85 $\mu$eV.

\begin{acknowledgments}
We thank T.\ Baart, E.\ Kawakami, Yu.~V.\ Nazarov, P.\ Scarlino and M.\ Shafiei for helpful discussions and R.~N.\ Schouten, B.\ v.d.\ Enden, J.\ Haanstra, and R.\ Roeleveld for technical support. This work is supported by the Stichting voor Fundamenteel Onderzoek der Materie (FOM), the Intelligence Advanced Research Projects Activity (IARPA) Multi-Qubit Coherent Operations (MQCO) Program, a European Research Council (ERC) Starting Grant, the Alexander von Humboldt Foundation, and the Swiss National Science Foundation.
\end{acknowledgments}


\begin{thebibliography}
\bibitem{footnote2} (i) The $T_2^*$ fitted from the time-dependent data varied strongly as a function of microwave power, (ii) it was impossible to find a set of values for $T_2^*$ which produced a good fit for the time-dependent data as well as for the measured line widths, (iii) the optimal fit we produced required a microwave signal at the gate $\sim 10^{-2}$ times weaker than expected.
\bibitem{footnote1} Precise determination of $\alpha_0$ is difficult since several parameters, such as the lever arm, are only known for d.c.\ signals.
\end{thebibliography}
\end{document}